# Revealing the Dominance of the Orbital Hall Effect over Spin in Transition Metal Heterostructures

J. L. Costa[1,*], E. Santos[1], J. B. S. Mendes[2], and A. Azevedo[1,†]

[1]*Departamento de Física, Universidade Federal de Pernambuco, Recife, Pernambuco 50670-901, Brazil.*
[2]*Departamento de Física, Universidade Federal de Viçosa, 36570-900 Viçosa, Minas Gerais, Brazil.*

We study inverse spin and orbital Hall effects in 19 transition metals using spin-pumping driven by ferromagnetic resonance. Spin-to-charge conversion was measured in YIG/X(5), while orbital-to-charge conversion was probed in YIG/Pt(2)/X(5) heterostructures. Here, X represents the different transition metals. Surprisingly, the orbital contribution overwhelmingly dominates over the spin response, clarifying the challenge of disentangling these effects. Our results largely agree with first-principles predictions for spin and orbital Hall conductivities but reveal discrepancies in select materials. These findings emphasize the fundamental role of the orbital Hall effect, and position orbitronics as a pivotal frontier in condensed matter physics.

## I. INTRODUCTION

Since the discovery of the spin Hall effect (SHE)[1-3], spin current manipulation has been widely explored[4-6]. Recently, attention has shifted to the electron orbital angular momentum (OAM) giving rise to the field of orbitronics[7,8]. The possibility of manipulating the OAM through electrical currents has opened new paths for magnetic switching, based on the orbital Hall effect (OHE)[8–10]. Unlike the SHE, the OHE is independent of spin-orbit coupling (SOC) and persists even with quenched OAM[9], making orbital effects more universal and stronger than spin effects, occurring in oxides, light metals, semiconductors, two-dimensional materials and in different types of magnetic materials[11-18]. This versatility is promising for applications like in the development of Magnetic Random Access Memory (MRAM)[19,20]. Notably, abundant materials, such as Ti and Cu, exhibit strong orbital response, enabling significant orbital torques without requiring strong SOC[12-14, 21, 22].

The inverse spin Hall effect (ISHE) and the inverse orbital Hall effect (IOHE) are reciprocal counterparts of the SHE and OHE, respectively, as dictated by the Onsager's reciprocity principle[23]. While the SHE converts charge currents into spin currents, the ISHE reverses this process. Similarly, the OHE generates orbital currents from charge current, whereas the IOHE converts orbital back into charge current. However, unlike SHE and ISHE, which exhibit strict reciprocity even at the local level, recent experiments reveal that the reciprocity between the OHE and IOHE can be violated locally, despite being preserved globally, as required by Onsager's principle. This local non-reciprocity arises from the non-conservation of orbital angular momentum under crystal fields, leading to distinct spatial profiles and even signs reversals in local orbital Hall conductivities between direct and inverse processes. Such findings reveal a fundamental distinction between spin and orbital transport phenomena, with important implications for orbitronics[24].

Direct observation of the OHE remains challenging, but spin pumping via ferromagnetic resonance (SP-FMR) enables its indirect detection[11,12,14-16,25]. In this method, FM/NM bilayers are used where the FM's magnetization dynamics and strong SOC inject spin or orbital currents into the NM layer. These currents are then converted to charge via ISHE and IOHE[14,26]. Recent experiments show an alternative mechanism for generation of orbital currents[11,12,14-16]. Under FMR, spin accumulation at the YIG/Pt interface diffuses as a spin current. The strong SOC of Pt converts this into a coupled spin-orbital current that partially transforms into charge via ISHE in Pt, while the remainder propagates to the NM layer for conversion through both ISHE and IOHE. The efficiency of these conversion processes - spin-to-charge and orbital-to-charge – is quantified by the respective spin Hall ($\theta_{SH}$) and orbital Hall ($\theta_{OH}$) angles, defined as $\theta_{(SH),OH} = \sigma_{(SH),OH}/\sigma_e$, where $\sigma_{(SH),OH}$ are the spin Hall conductivities (SHC) or orbital Hall conductivities (OHC), and $\sigma_e$ is the electrical conductivity of the NM. While early tight-binding and first principles calculations predicted exclusively positive $\sigma_{OH}$[10,28,29], recent Wannier-interpolation studies reveal negative $\sigma_{OH}$ values for certain materials or material phases[30], challenging earlier theoretical expectations.

In this work, we present SP-FMR studies on two heterostructures series. Series A: YIG/X and series B: YIG/Pt(2)/X(5), where X represents transition metals (numbers denote layer thicknesses in nm), and YIG ($Y_3Fe_5O_{12}$) stands for Yttrium Iron Garnet. While the transition metal layers were fabricated by DC magnetron sputtering, YIG is grown by Liquid Phase Epitaxy (LPE). The YIG/Pt(2) bilayer serves as an orbital current injector, enabling IOHE detection in the X layer. ISHE and IOHE measurements employed FMR-generated spin currents (9.4 GHz microwave cavity, TE₁₀₂ mode, Q ≈ 2000).

[*]Email: jefferson.limacosta@ufpe.br
[†]Email: antonio.azevedo@ufpe.br

## II. SPIN AND ORBITAL HALL EFFECTS

Following Tserkovnyak *et al.*'s theory[31], the precessing magnetization of a ferromagnetic layer can pump spin currents into adjacent nonmagnetic materials. In materials with strong SOC, this spin current converts to charge current via the inverse SHE. For bulk-mediated conversion, the ISHE-induced charge current density is[32]:

$$J_c^{ISHE} = g_{eff}^{\uparrow\downarrow} \lambda_s \theta_{SH} f\left(\frac{e}{2\hbar}\right) L p_{xz} \tanh\left(\frac{t}{2\lambda_s}\right) L(H - H_r) \cos\phi, \quad (1)$$

where $h_{rf}$ is the amplitude of microwave magnetic field with frequency $f$, $t$ is the film thickness, $L$ is the sample width, $g_{eff}^{\uparrow\downarrow}$ represents the effective spin mixing conductance that quantifies how efficiently spin current is transferred across the FM/NM interface. $\lambda_s$ is the spin diffusion length in the normal metal, and $p_{yz}$ expresses the ellipticity and the spatial variation of the RF magnetization of the FMR mode. $L(H - H_r)$ is a lorentzian function with parameters $H_r$ (resonance field) and $\Delta H$ (resonance linewidth). The angle $\phi$, which defines the orientation of the applied magnetic field relative to the electrical contacts, is illustrated in Fig. 1(a). The detected voltage shows a Lorentzian lineshape centered at $H = H_r$, as shown in Fig. 1(b) for YIG/Pt(2). In materials with strong SOC, the injected spin current $\vec{J}_s$ can additionally generate orbital current $\vec{J}_L$, leading to a coupled spin-orbital current $\vec{J}_{LS}$.

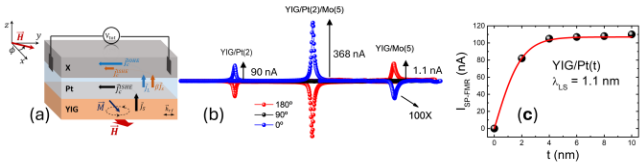

**Fig. 1.** Experimental scheme used for SP-FMR signal measurements. (a) Under resonance conditions, a spin current $\vec{J}_s$ is injected into the Pt layer, and the strong SOC induces a coupled current $\vec{J}_{LS}$. The orbital component that reaches the top metal layer is converted into charge current via the IOHE. (b) Typical SP-FMR curves for YIG/Pt, YIG/Pt(2)/Mo(5), and YIG/Mo(5). (c) SP-FMR signals for YIG/Pt($t_{Pt}$) as a function of Pt thickness.

As illustrated in Fig. 1(a), the total voltage $V_{tot}$ measured in YIG/Pt($t_{Pt}$)/X heterostructure contains two components: a fixed contribution $V_{ISHE}^{Pt(2)}$ from the Pt(2) layer (determined via separate YIG/Pt(2) measurement), and a material-dependent contribution $V_{sp}^{X(5)}$ from the X layer that combines ISHE and IOHE effects. Accordingly, the total signal can be expressed as:

$$V_{tot} = V_{ISHE}^{Pt(2)} + \left(V_{IOHE}^{X(5)} + \beta V_{ISHE}^{X(5)}\right), \quad (2)$$

where $\beta$ (with $\beta < 1$) represents the fraction of the spin current transmitted across the Pt(2)/X interface. From Fig. 1(c), for $t_{Pt}$ = 2.0 nm, the spin current reaches 23% of its saturation value, indicating that only 23% of the spin current injected at the YIG/Pt(2) interface reaches the Pt(2)/X(5) interface. Thus, $\beta \approx 0.23$. The charge current density generated via the ISHE is described by $\vec{J}_C^{ISHE} = (\hbar/2e)\,\theta_{SH}(\vec{J}_S \times \hat{\sigma}_S)$, while analogously, the IOHE produces $\vec{J}_C^{IOHE} = (\hbar/2e)\,\theta_{OH}(\vec{J}_L \times \hat{\sigma}_L)$, where $\theta_{SH}$ and $\theta_{OH}$ are the spin and orbital Hall angles, and $\hat{\sigma}_S$ and $\hat{\sigma}_L$ are the corresponding polarization directions. The spin current reaching the Pt/X interface depends on the Pt layer thickness, as shown in Fig. 1(c). As $t_{Pt}$ increases, the charge current saturates at 109 nA. Consequently, the ISHE contribution from the X layer in Eq. (2) is scaled by $0.23 \times V_{ISHE}^{X(5)}$, where $V_{ISHE}^{X(5)}$ corresponds to the SP-FMR signal measured in the reference YIG/X(5) sample under the same RF excitation. The SP-FMR attributed to the IOHE in the X layer is thus extracted as: $V_{IOHE}^{X(5)} = V_{tot} - V_{ISHE}^{Pt(2)} - 0.23 \times V_{ISHE}^{X(5)}$. For example, the middle signal in Fig. 1(b) (368 nA) corresponds to the total SP-FMR signal measured in the YIG/Pt(2)/Mo(5) heterostructures. The signals are expressed in nA by dividing the measured voltage by the electric resistance between the electrodes ($I_{tot} = V_{tot}/R$). The left and right signals correspond the ISHE contributions from the YIG/Pt(2) and YIG/Mo(5) bilayers, respectively: $I_{ISHE}^{Pt(2)} = 90$ nA and $I_{ISHE}^{Mo(5)} = -1.1$ nA. Substituting these values: $I_{IOHE}^{X(5)} = 368 - 90 - 0.23 \times (-1.1) \cong 278$ nA. In this case, the ISHE contribution from the Mo layer is negligible, allowing a clean estimation of the IOHE. However, for materials with strong SOC – such as W, Ir, or Pd - the ISHE contribution of the X material is significant and must be explicitly accounted.

Fig. 2 shows the SP-FMR results for the 19 analyzed materials in sample series A and B. The inset displays the corresponding ISHE signals for series B. The experimental data were fitted using Lorentzian lineshapes, as described by Eq. (1). For YIG/X(5) (series A), the measured signal arises from the ISHE in the X layer, as shown in the insets of Fig. 2. In YIG/Pt(2)/X(5) (series B) structures, the signal includes contributions from both the ISHE in YIG/Pt(2) and the combined ISHE and IOHE from the X(5) layer. The SP-FMR signals observed in series A are well explained by Eq. (1). The signal polarity reflects the sign of $\sigma_{SH}$: at $\phi = 0°$ (blue data), the polarity appears as expected, while reversing the magnetic field $\vec{H}$ to $\phi = 180°$ (red data) flips the spin polarization ($\hat{\sigma}_s$), thereby inverting the ISHE polarity, typically with only minor changes in amplitude. At $\phi = 90°$ (black data), the signal vanishes, consistent with the expected angular dependence. Samples in series B exhibit identical behavior, which is attributed to the strong SOC in Pt, which couples (**S**) and orbital angular momentum (**L**). Consequently, reversing $\vec{H}$ inverts both ISHE and IOHE signals. All SP-FMR measurements were performed at fixed RF power of 7.3 mW, and all samples were cut from the same YIG wafer to ensure consistency. By applying Eq. (2) to each sample, the contribution of the IOHE can be extracted. It is noteworthy that YIG/Pt(2nm)/Cu shows no signal increase compared to YIG/Pt(2nm), which demonstrates negligible IOHE and ISHE in Cu films, as reported in Ref. [11].

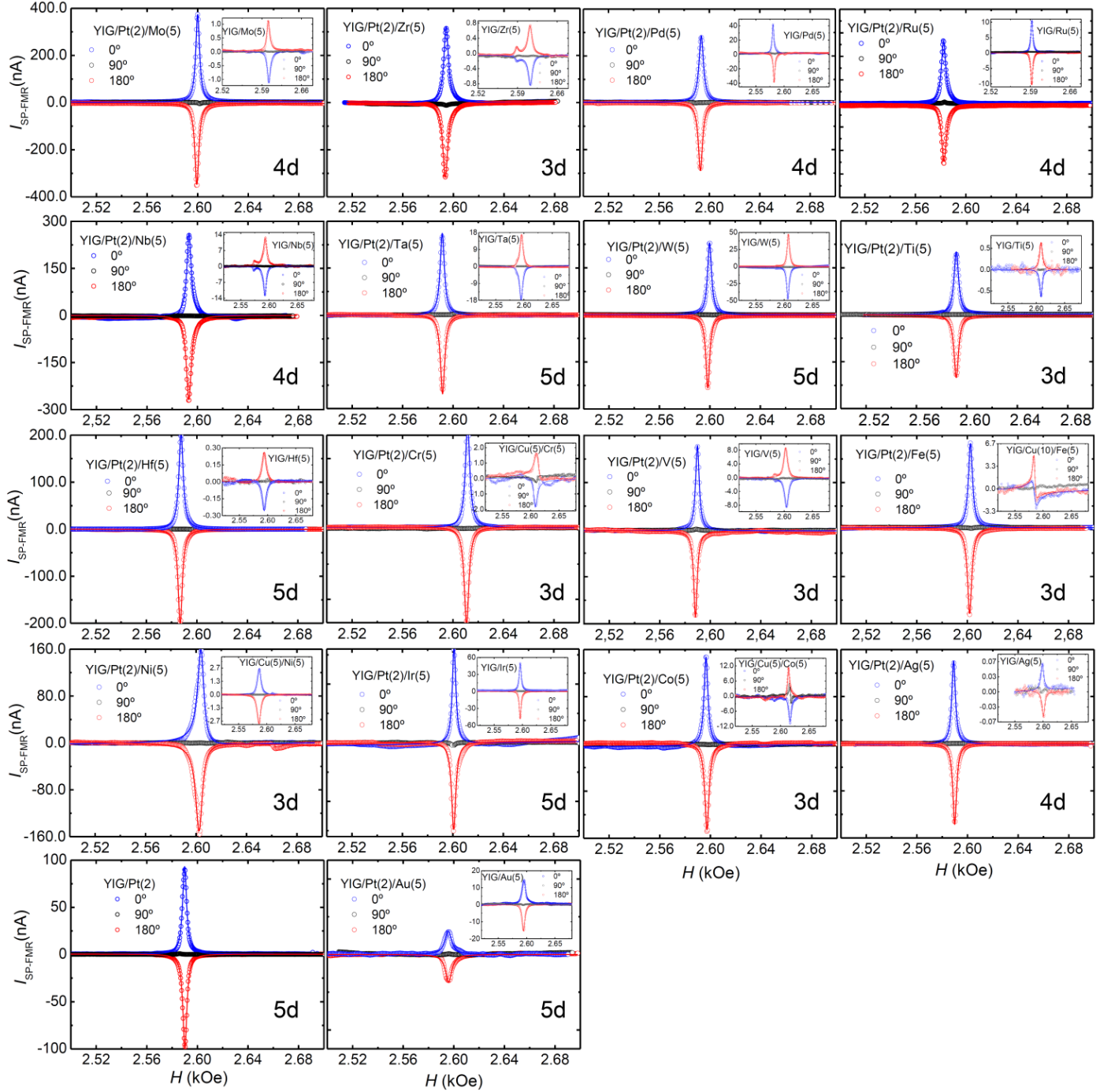

**Fig. 2.** SP-FMR signals for the three relevant angular configurations: blue data ($\phi = 0°$), red ($\phi = 180°$), and black ($\phi = 90°$). Each graph shows the total charge current due to the ISHE and the IOHE in YIG/Pt(2)/X(5) structures, while the inset highlights the electric current generated specifically by the ISHE at the YIG/X(5 nm) samples. The data can be fitted with a Lorentzian function of the form $I = I_S \Delta H^2 / [(H - H_r)^2 + \Delta H^2]$. We add a spacer layer of Cu (5nm) between the YIG (ferrimagnetic) and the magnetic materials Fe, Co, Ni and Cr (in its antiferromagnetic phase) to avoid magnetic coupling.

## III. DISCUSSION OF RESULTS

Fig. 3(a) shows the spin Hall conductivity values calculated from first principles by D. Go et al[30]. Fig. 3(b) shows the experimentally measured charge current arising from ISHE in YIG/X(5), fitted using the Lorentzian lineshape described by Eq. (1). The values plotted in Fig. 3(b) represent the average magnitudes of measurements taken at $\phi_H = 0°$ and $\phi_H = 180°$ for groups 3d, 4d and 5d. Fig. 3(c) presents the orbital Hall conductivity values, also from D. Go et al[30], while Fig. 3(d) shows the experimentally extracted charge current associated with the IOHE in the X layer in YIG/Pt(2)/X(5), obtained using Eq. 2. Notably, the correlation between theoretical predictions and experimental results is stronger for the ISHE than for the IOHE. This discrepancy highlights the greater sensitivity of orbital-related mechanisms to experimental factors such as crystalline texture, structural disorder, and interfacial coupling[24]. It is crucial to recognize that both the magnitude and sign of $\sigma_{OH}$ and $\sigma_{SH}$ depend strongly on the detailed electronic structure, particularly near the Fermi level. Since theoretical calculations are typically performed at ideal conditions, a more accurate comparison with experiments should account for real band structure under ambient conditions, especially in polycrystalline materials.

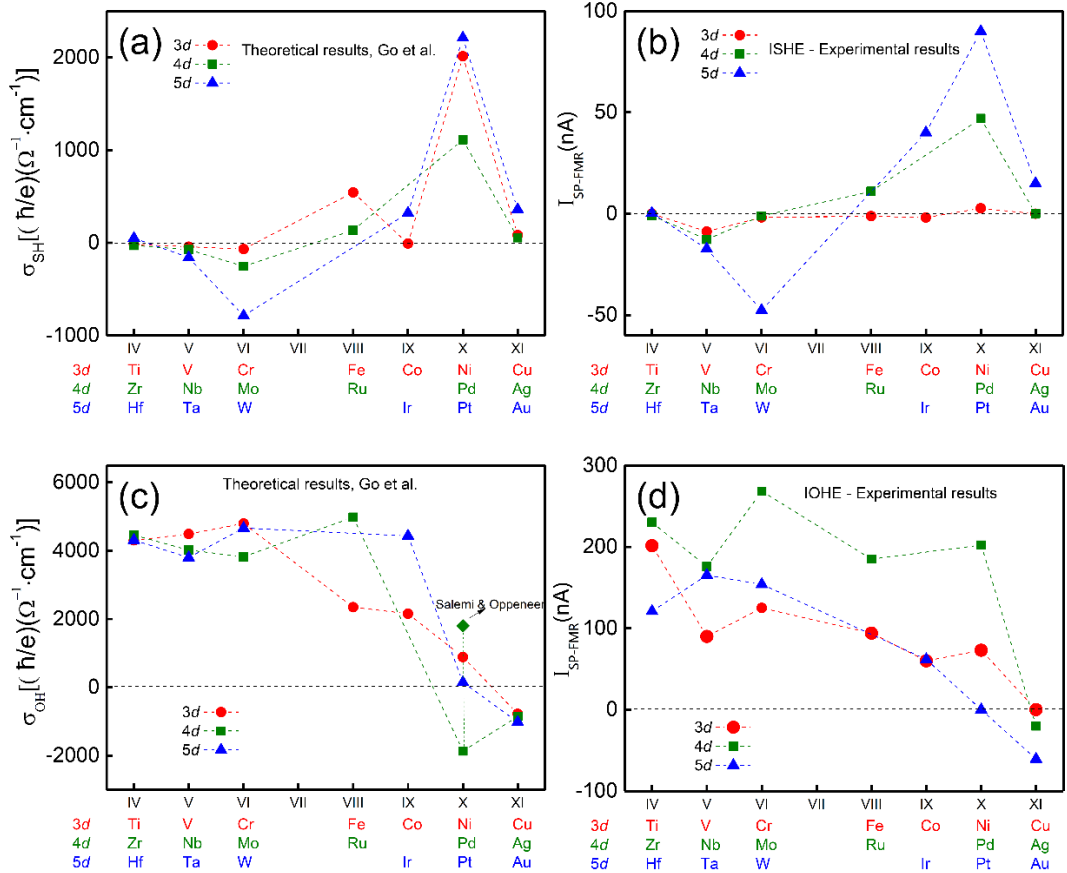

**Fig. 3.** (a) and (c) calculated results for the $\sigma_{SH}$ and $\sigma_{OH}$ via Wannier interpolation, adapted from Ref. [28, 30]. (b) and (d) experimentally measured values of the ISHE and IOHE. It is important to note that the vertical scale of the IOHE is three times larger than that of ISHE.

We present a quantitative comparison of all investigated elements, starting with the 3d transition metals. For clarity, we label the heterostructures YIG/X(5) and YIG/Pt(2)/X(5) as series A and B, respectively.

*3d transition metals.* For the samples in series A, it is important to highlight that the 3d metals Fe, Co, Ni and Cr are in their magnetic phases. In particular, the antiferromagnetic phase of Cr[33] is identified by the asymmetric spin-pumping signal characteristic of magnetic coupling (see inset for Cr in Fig. 2). To mitigate interfacial exchange interactions between magnetic layers, a Cu(5) spacer layer was introduced. Comparing the theoretical and experimental results for the SHC and the ISHE, shown in Figs. 3(a) and 3(b), respectively, we observe a general agreement in trends with exceptions for Fe and Ni. While theoretical studies have predicted a positive theoretical SHC for Fe[28-30], our experiment reveal a negative value. As discussed in Ref. [34], Fe in its magnetic phase can exhibit positive or negative SHC. A similar behavior is found for Ni. Our results show a weak ISHE signal, indicating a low spin-to-charge current conversion efficiency, comparable in magnitude to that observed for Cr. However, our results for the $\theta_{SH}$ of Ni and Cr agree very well with the theoretical results[30]. The low values observed for both Ni and Fe can be attributed to the strong dependence of SHC on the Fermi level[34]. In contrast, our results for Co and Ti exhibit excellent agreement with numerical prediction[30]. In the case of Cu, we adopted previously published results obtained under similar experimental conditions and parameters[11], which confirms that Cu has negligible SHC and OHC.

For the samples in series B, the theoretical and experimental results show a consistent trend (Figs. 3 (c) and 3 (d)). While the calculated OHC for Ti, V and Cr are nearly identical, the experimentally measured IOHE values shows noticeable variations. Previous work[11] confirmed that Cu exhibits negligible IOHE, consistent with the theoretically predicted negligible OHC[30]. However, Cu can exhibit a positive OHC depending on the Fermi energy[30]. Their analysis reveals that negative OHC values only emerge when the anomalous position correction to the velocity operator is included. While this correction has minimal impact on spin-related effects in most materials, it is essential for accurately capturing orbital-mediated transport phenomena.

*4d transition metals.* When analyzing the samples of series A (YIG/X(5), with X being a 4*d* transition metal), we observed significant variations in the ISHE signals depending on the material. For Zr, Nb, Mo, and Ru, the signal intensity remained below 13 nA, a value significantly lower than that of the reference sample YIG/Pt(2). The measured SHC values were consistent with theoretical predictions[30]: positive for Ru, and negative for Zr, Nb, and Mo. Ag, on the other hand, exhibited virtually no ISHE signal, a result that aligns well with its weak spin–orbit coupling and low density of d-states at the Fermi level.

By examining the experimental results in series B for 4*d* transition metals, we found surprisingly intense signals in elements not traditionally associated with strong SOC. Mo, for example, showed an orbital signal of over 260 nA, surpassing not only the other 4*d* elements but also many 5*d* ones, providing clear evidence of orbital effects. This suggests that Mo combines high electrical conductivity with strong orbital response, making it a lighter and more abundant alternative for applications that previously focused on noble metals. Another highlighted material is Ru, which presents a strong signal, although smaller than that of Mo. The difference between the two may be less related to intrinsic SOC and more to how the electronic structure and crystal symmetry influence orbital angular momentum transport. The contrast between them likely reflects not just intrinsic SOC strength, but also the influence of electronic structure and crystal symmetry on orbital angular momentum transport. These findings point to a key distinction: orbital transport appears significantly more sensitive to structural and experimental conditions than spin transport, which may help explain discrepancies between theory and experiment. In the case of Pd, we found a large and positive IOHE signal, contradicting the results of D. Go et al [30], but in agreement with the results of Salemi and Oppeneer[28], suggesting the need for further theoretical refinement. Completing the analysis, Zr and Nb showed quite high intensities: 230 nA for Zr and 176 nA for Nb. Furthermore, the fact that Nb and Zr exhibit both high IOHE and negative (or nearly zero) negative SHC is a strong indication that orbital angular momentum transport can occur even in the absence of significant ISHE, reinforcing the idea that IOHE and ISHE are partially independent mechanisms, sensitive to different material properties. This can be extremely useful for isolating and studying IOHE in systems where ISHE is not dominant. Finally, regarding Ag, we found a negative IOHE, associated with a reduction in the SP-FMR signal in YIG/Pt(2) from 90 nA to 60 nA, i.e. IOHE signal of -30 nA. This result confirms that Ag presents a considerable OHC, in agreement with[30].

*5d transition metals.* Materials exhibiting the strongest spin-related effects generically distinguish the 5d series, as consistently confirmed by the experimental observations. The intense SOC, characteristic of these elements, strongly favors spin-to-charge conversion via ISHE. This is particularly evident in metals such as Pt, Ir, W, and Ta. Pt remains the element with the strongest ISHE, presenting a robust signal of approximately 90 nA (for the experimental parameters used in this investigations), consistent with a high and positive SHC, widely reported in the literature[13,28-30,34,35]. In contrast, W and Ta show negative ISHE signals of relatively large magnitude. This behavior is characteristic of metals situated on the left side of the 5*d* block, where the *d*-band is less than half-filled, a condition that, according to theoretical predictions, leads to negative SHC[10]. Furthermore, both W and Ta are in their $\beta$-phase, promoted by the small thickness, only 5 nm, a phase known for its enhanced SHC[36,37]. Ir, in turn, exhibits a positive and well-defined ISHE signal of around 40 nA (for the experimental parameters used in this investigations), confirming its role as one of the most balanced 5d metals in terms of spintronic response.

By analyzing the total signals obtained in series B, both spin (ISHE) and orbital (IOHE) contributions are evident. As mentioned previously, these components are extracted by comparing to the reference sample YIG/Pt(2). Interestingly, while Pt is dominant in ISHE, its IOHE contribution is practically zero in our experiments - a result consistent with prior studies[30]. However, techniques such as ST-FMR[38] and Terahertz[39] spectroscopy have indicated that Pt may exhibit a non-negligible IOHE under specific conditions. This reinforces the fact that IOHE is a more subtle effect and highly sensitive to the technique used and to the local electronic and crystalline structure. For W and Ta, the total signals remain significant, yet are slightly lower than those measured for 4d metals like Mo or Zr, suggesting that although high SOC favors both spin and orbital effects, orbital conversion efficiently does not scale straightforwardly with atomic number Z. Instead, it depends on orbital band texture, Fermi level positioning, and crystal symmetry. Ir represents an intermediate case, exhibiting a strong IOHE signal of 150 nA (for the experimental parameters used in this investigations). This, combined with its relatively high ISHE, makes Ir one of the few known materials capable of efficiently transporting both spin and orbital angular momentum. Finally, Au presents an atypical behavior: it exhibits a negative IOHE of -60 nA, in agreement with recent theoretical predictions[30]. This highlights the diversity of orbital responses among 5d metals and reinforces the notion that IOHE is strongly influenced by $d$-band filling and band structure details.

Linewidth broadening ($\delta H$) plays an important role in accurately attributing signal contributions to ISHE and IOHE. To properly account for these effects, we systematically measured linewidth changes across all samples. Fig. 4 shows $\delta H$, defined as the difference between the linewidth $\Delta H$ measured in YIG/X(5) or YIG/Pt(2)/X(5), and the intrinsic YIG linewidth ($\Delta H_0$). We observe that the linewidth broadening in YIG/Pt(2)/X(5) (Fig. 4(a)) exhibits lower dispersion compared to YIG/X(5) (Fig. 4 (b)), indicating that the IOHE does not significantly cause linewidth broadening in these heterostructures. For YIG/Pt(2)/X(5), the most pronounced linewidth increase occurs with Ni, though this is likely caused by magnetic coupling rather than spin-pumping-induced damping. We also note that light elements like Ag, Zr and Cu induce negligible additional damping due to their low spin-pumping efficiency, while heavy elements like Pt, Pd and W introduces significant damping via spin pumping (Fig. 4(b)).

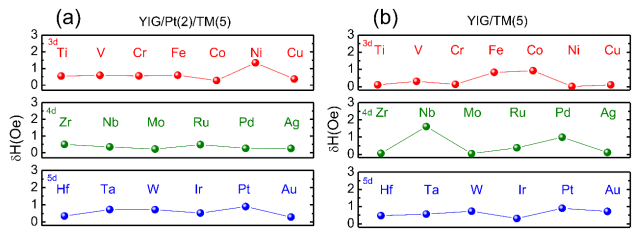

**Fig. 4.** Linewidth broadening $\delta H$. (a), $\delta H$ due to spin-pumping in YIG/Pt(2)/X(5). (b), $\delta H$ due to spin-pumping in YIG/X(5).

## V. CONCLUSIONS

In this work, we performed a systematic and comprehensive investigation of the inverse spin and orbital Hall effects across 3d, 4d, and 5d transition metals, using spin pumping technique via ferromagnetic resonance as a precise and versatile probe. Our experimental results show strong consistency with theoretical predictions for spin Hall conductivity (SHC) and orbital Hall conductivity (OHC), particularly those based on first-principles calculations. We observed that materials like Au and Ag are better described by first-principles approach than by tight-binding models, especially regarding their SHC and OHC. Our data reveal clear trends in the spintronic and orbitronic responses across the transition metals series. In most cases, the orbital contribution to the generated charge current exceeds the spin contribution, highlighting the growing relevance of orbitronics as an independent as essential field within spintronics. Importantly, our results challenge the prevailing assumption that only heavy metals with strong SOC are efficient in generating charge currents via spin pumping. We observed that even metals with relatively weak SOC can exhibit a strong IOHE, highlighting the influence of orbital band structure and interfacial properties. Furthermore, we provide experimental confirmation of negative OHC in certain transition metals – a result that resolves longstanding ambiguities in the literature, where early studies predominantly reported only positive values. The comparison between experimental data and theoretical models reinforces the notion that ISHE and IOHE are fundamentally distinct physical effects, each sensitive to different structural and electronic properties parameters. In particular, IOHE appears more susceptible to structural imperfections, disorder, orbital texture, and interfacial conditions, emphasizing the need for detailed material characterization. Finally, our investigation establishes spin-pumping FMR as a highly sensitive and accessible platform for the detection and quantification of orbital currents, thereby opening a new experimental avenue for the systematic exploration of IOHE across a broad range of material systems. By validating key theoretical predictions and identifying promising candidates for orbitronic functionality, our work not only advances the fundamental understanding of orbital transport but also lays a solid foundation for the development of next-generation spin-orbitronic technologies that exploit both spin and orbital degrees of freedom.

## ACKNOWLEDGMENTS

This research is supported by Conselho Nacional de Desenvolvimento Científico e Tecnológico (CNPq), Coordenação de Aperfeiçoamento de Pessoal de Nível Superior (CAPES) (Grant No. 0041/2022), Financiadora de Estudos e Projetos (FINEP), Fundação de Amparo `a Ciência e Tecnologia do Estado de Pernambuco (FACEPE), Universidade Federal de Pernambuco, Multiuser Laboratory

## DATA AVAILABILITY

The data that support the findings of this study are available from the corresponding author upon reasonable request.

---

Figure 1

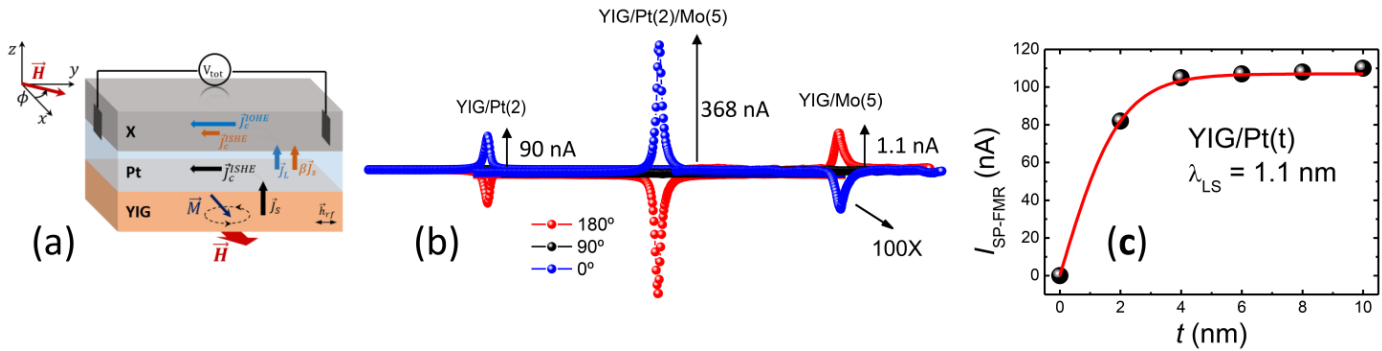

Figure 2

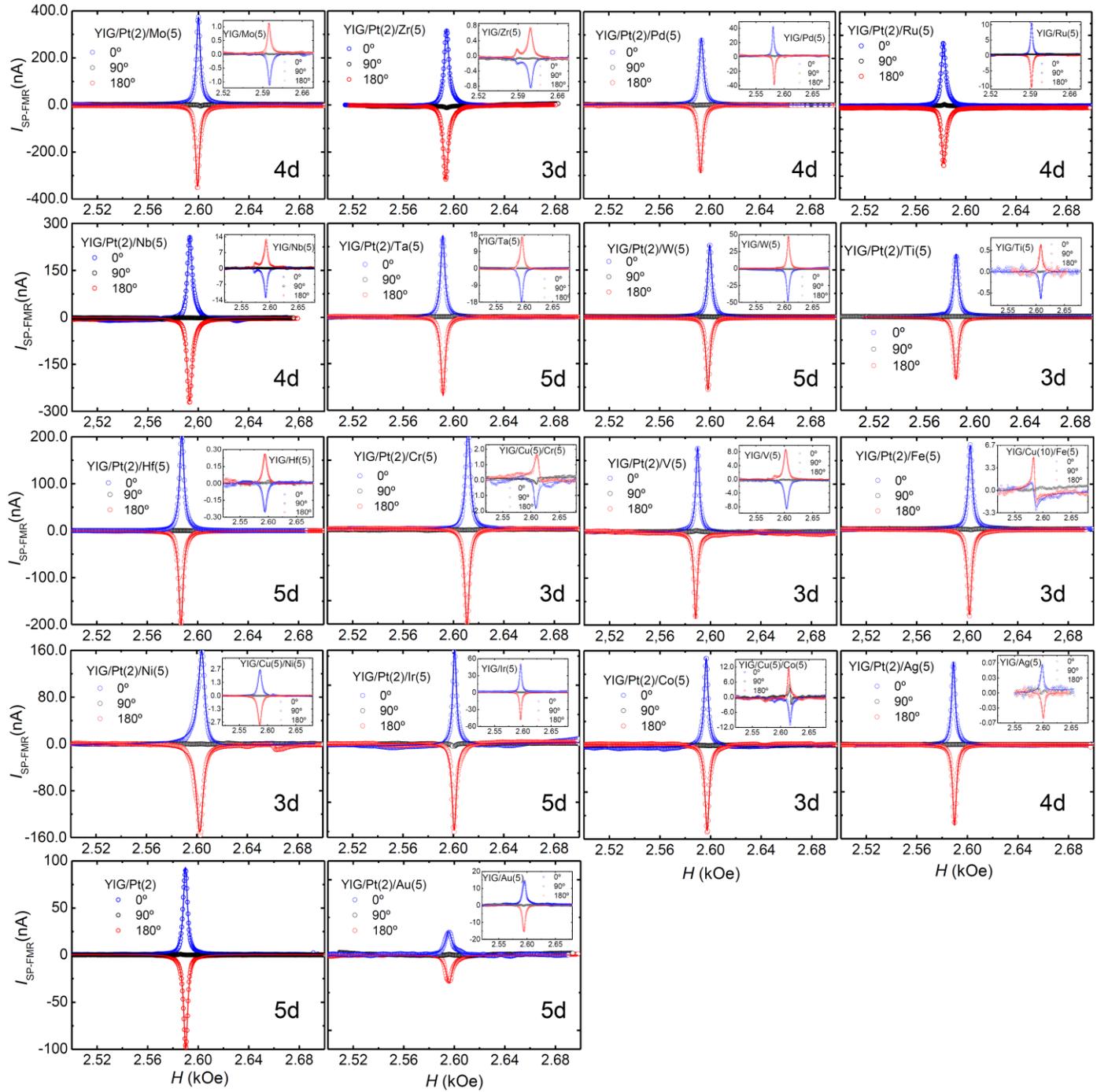

Figure 3

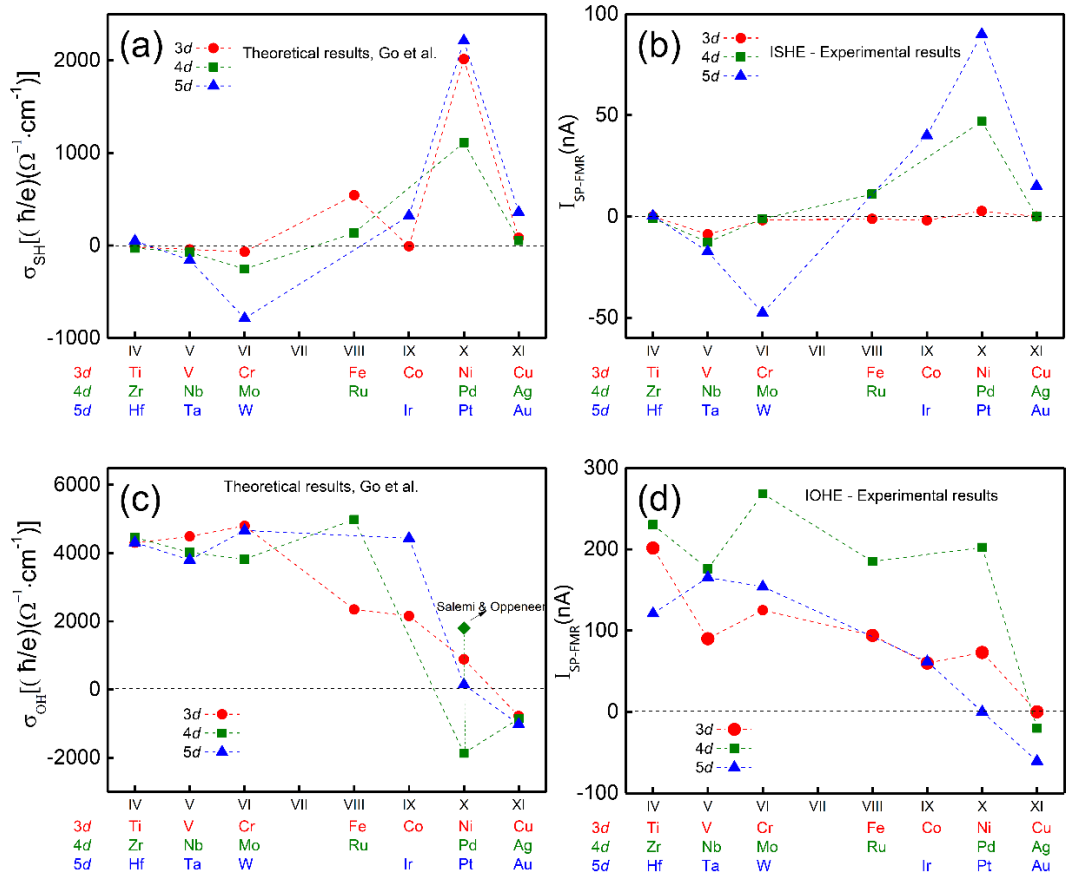

Figure 4

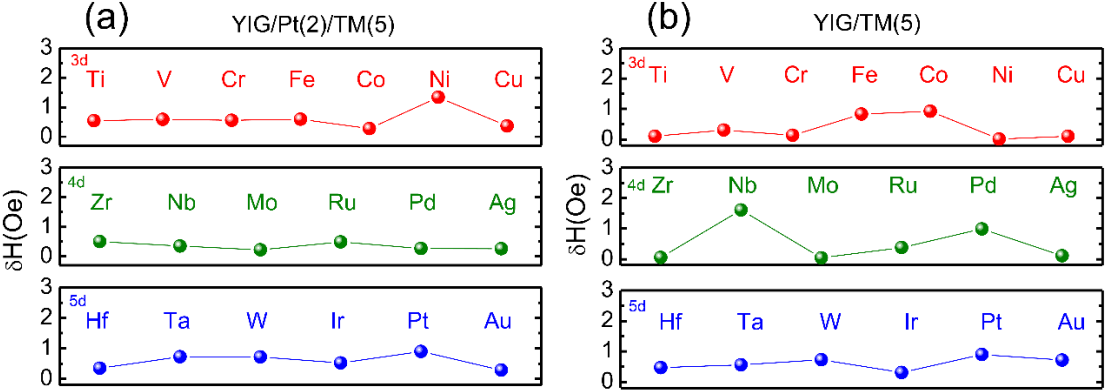